\documentclass[aps, pra, twocolumn, superscriptaddress]{revtex4-1} 
\usepackage{amssymb, amsmath, amsthm}
\usepackage{xcolor}
\usepackage{graphicx}
\usepackage{hyperref}
\usepackage{comment}
\usepackage{blkarray}
\usepackage{mathtools}
\usepackage{tikz}
\usetikzlibrary{decorations.pathreplacing}
\usepackage{dsfont}
\usepackage{braket}

\usepackage{graphicx}
\usepackage{dcolumn}
\usepackage{bm}

\newcommand{\Tr}[1]{\mathrm{Tr}}

\def\tU{\tilde{U}}

\newcommand{\titleinfo}{
Generalized Deep Thermalization for Free Fermions}
\begin{document}

\preprint{APS/123-QED}

\title{\titleinfo
}

\author{Maxime Lucas}
\affiliation{Laboratoire de Physique Th\'eorique et Mod\'elisation, CNRS UMR 8089,
	CY Cergy Paris Universit\'e, 95302 Cergy-Pontoise Cedex, France}

\author{Lorenzo Piroli}
\affiliation{Philippe Meyer Institute, Physics Department, \'Ecole Normale Sup\'erieure (ENS), Universit\'e PSL, 24 rue Lhomond, F-75231 Paris, France}

\author{Jacopo De Nardis}
\affiliation{Laboratoire de Physique Th\'eorique et Mod\'elisation, CNRS UMR 8089,
	CY Cergy Paris Universit\'e, 95302 Cergy-Pontoise Cedex, France}

\author{Andrea De Luca}
\affiliation{Laboratoire de Physique Th\'eorique et Mod\'elisation, CNRS UMR 8089,
	CY Cergy Paris Universit\'e, 95302 Cergy-Pontoise Cedex, France}

\date{\today}

\begin{abstract}
	In non-interacting isolated quantum systems out of equilibrium, local subsystems typically relax to non-thermal stationary states. In the standard framework, information on the rest of the system is discarded, and such states are described by a Generalized Gibbs Ensemble (GGE), maximizing the entropy while respecting the constraints imposed by the local conservation laws. Here we show that the latter also completely characterize a recently introduced projected ensemble (PE), constructed by performing projective measurements on the rest of the system and recording the outcomes. By focusing on the time evolution of fermionic Gaussian states in a tight-binding chain, we put forward a random ensemble constructed out of the local conservation laws, which we call deep GGE (dGGE). For infinite-temperature initial states, we show that the dGGE coincides with a universal Haar random ensemble on the manifold of Gaussian states. For both infinite and finite temperatures, we use a Monte Carlo approach to test numerically the predictions of the dGGE against the PE. We study in particular the $k$-moments of the state covariance matrix and the entanglement entropy, finding excellent agreement. Our work provides a first step towards a systematic characterization of projected ensembles beyond the case of chaotic systems and infinite temperatures.
\end{abstract}

\maketitle

\section{Introduction}
 The established paradigm for quantum thermalization in isolated quantum systems is extremely simple, and yet surprisingly effective. When a system is initialized in a short-range correlated state, it predicts, under a few typicality assumptions, that the late-time properties of local subsystems are described by a thermal Gibbs ensemble~\cite{cazalilla2010focus,polkovnikov2011colloquium,dalessio2016quantum}. In this framework, usually understood in terms of the eigenstate thermalization hypothesis~\cite{deutsch1991,srednicki1994chaos,rigol2008thermalization}, 
one is interested in a local subsystem, while its complement plays the role of a thermal bath which is assumed not to be observed (\emph{i.e.} measured).

Thermalization and its mechanisms have been probed to exquisite detail in a number of cold-atomic experiments~\cite{trotzky2012probing,langen2013local,geiger2014local,langen2015experimental,neill2016ergodic,clos2016time,kaufman2016quantum}. In fact, these works fully demonstrate the ability of current setups to keep track of both subsystems and their complement, having access to information on the ``bath'' which is discarded in the traditional setting. Motivated by this, two recent works~\cite{cotler2021emergentquantum,choi2021emergentrandomness} have put forward a new perspective, in which one is interested in the ensemble describing a subsystem $A$ when its complement, $B$, is observed via projective measurements. This gives rise to an ensemble of pure states in $A$, called \emph{projective ensemble} (PE), which can be thought of as a particular unraveling of the subsystem density matrix.

Based on numerical and experimental evidence, Refs.~\cite{cotler2021emergentquantum,choi2021emergentrandomness} found that, for chaotic dynamics and infinite-temperature initial states, the PE approaches a Haar-random ensemble over the set of pure states in $A$, forming a \emph{quantum state design}~\cite{renes2004symmetric,ambainis2007quantum}. From the fundamental standpoint, the appeal of this result lies in its universality, as it is claimed to be independent of any microscopic detail. Subsequent work substantiated these findings, with rigorous results provided in Refs.~\cite{ho2022exact,claeys2022emergent,ippoliti2022dynamical} for classes of chaotic \emph{dual-unitary} quantum circuits~\cite{bertini2019entanglement,bertini2019exact,piroli2020exact}, while further connections between the onset of thermalization and quantum state designs were investigated in Refs.~\cite{ wilming2022high,ippoliti2022solvable}.

\begin{figure}
	\includegraphics[scale=0.5]{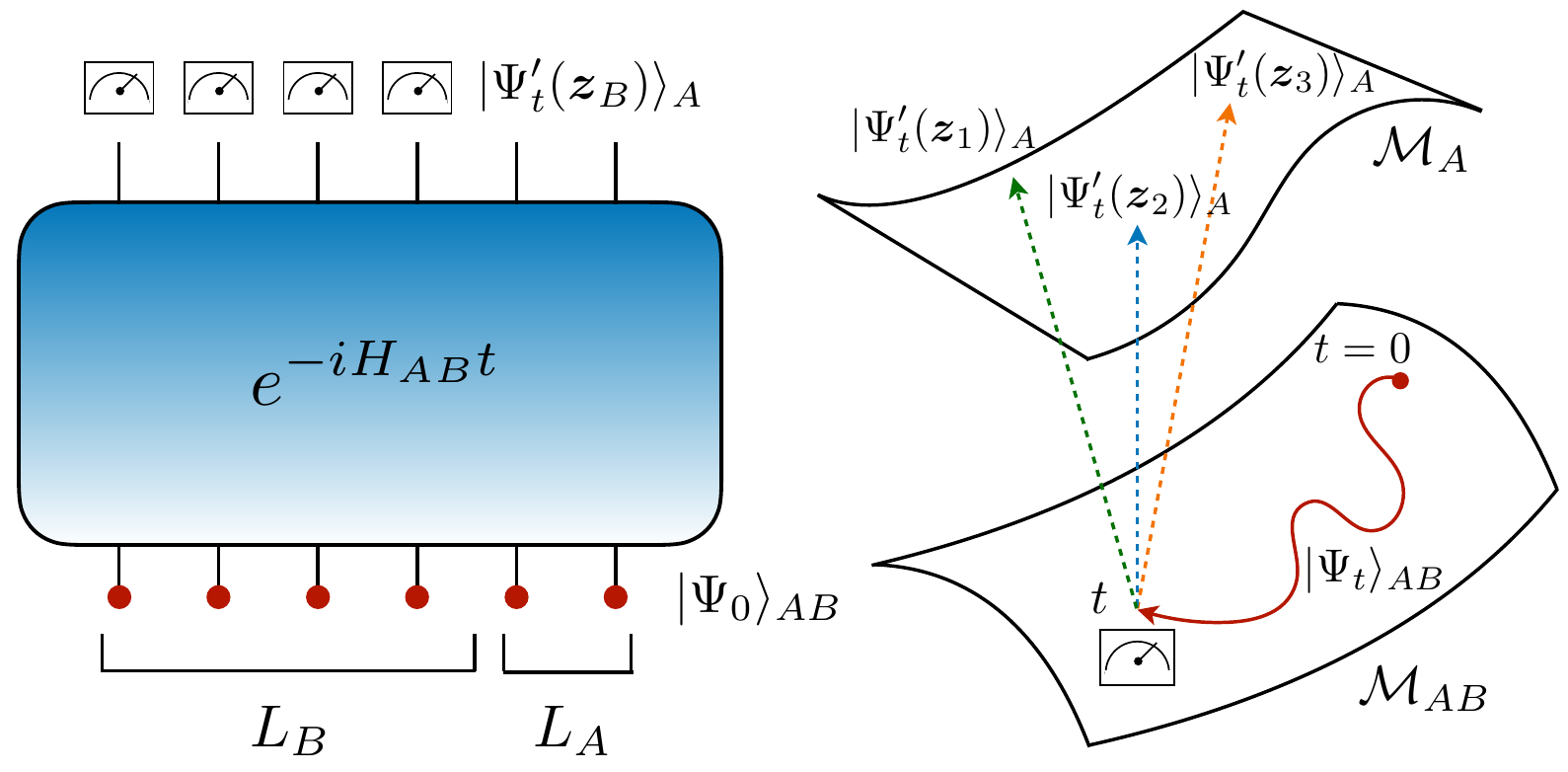}
	\caption{An initial Gaussian state $\ket{\Psi_0}_{AB}$ evolves according to a quadratic Hamiltonian $H$. The unitary dynamics remains within the manifold of Gaussian states $\mathcal{M}_{AB}$. After the measurement, the reduced state of the system is projected onto a pure Gaussian state $\ket{\Psi^\prime_t(\boldsymbol{z})}_{A}\in \mathcal{M}_A$ depending on the outcomes $\boldsymbol{z}$. We are interested in the ensuing ensemble on $\mathcal{M}_A$. }
	\label{fig:sketch}
\end{figure}

A natural question is how this picture is modified in the presence of conservation laws, including in particular \emph{integrable} systems \cite{korepin1997quantum,essler2005one,takahashi2005thermodynamics}, nowadays easily realized experimentally~\cite{Malvania_2021,langen2015experimental,Wang_2022}. In this case, local subsystems approach a stationary state described by a generalized Gibbs ensemble (GGE)~\cite{rigol2007relaxation}, built out of all the quasi-local conserved operators (or \emph{charges})~\cite{ilievski2015complete,ilievski2016string,vidmar2016generalized,essler2016quench,ilievski2016quasilocal,piroli2016exact,piroli2017correlations,ilievski2017from,pozsgay2017generalized}. GGEs are interesting as they differ qualitatively from thermal states, representing non-equilibrium phases with possibly exotic features~\cite{denardis2014solution,wouters2014quenching,pozsgay2014correlations,piroli2016mutiparticle,calabrese2016introduction}. Accordingly, one can ask how local conservation laws affect the PE. 

In this work, we tackle this problem in the simplest case of non-interacting systems. Focusing on the time evolution of fermionic Gaussian states in a tight-binding model, we put forward a random ensemble constructed out of the conserved charges, which we call \emph{deep} GGE (dGGE), and provide evidence of its validity based on Monte Carlo computations. For infinite-temperature initial states, we show that the dGGE coincides with a universal Haar random ensemble on the manifold of Gaussian states, while, for generic initial states, we introduce a generalized Haar ensemble to account for the finite expectation values of the charges.   

	The rest of this work is organized as follows. In Sec.~\ref{sec:model} we introduce the model we study. We also briefly recall the standard GGE and the PE. In Sec.~\ref{sec:deepGGE} we put forward our general conjecture for the deep GGE, while Sec.~\ref{sec:infinite_temp} shows how the latter leads to a universal ensemble for infinite-temperature initial states. Finally, our conclusions are consigned to Sec.~\ref{sec:conclusions}, while several appendices provide details on the most technical parts of our work.

\section{The model} 
\label{sec:model}
We consider a chain of spinless fermions, described by the Hamiltonian
\begin{equation}\label{eq:Ham}
	H= -\sum_{j=1}^L \left[c^{\dagger}_{j+1}c_{j}+c^{\dagger}_{j}c_{j+1}\right]\,,
\end{equation}
where $c_j$,  $c^\dagger_j$ are canonical operators satisfying $\{ c_i , c^\dagger_j\} = \delta_{i,j}$. We initialize the system in a short-range correlated state $\ket{\Psi_0}$ and consider a bipartition into a region $A$ and its complement $B$, ``the bath'', containing $L_A$ and $L_B$ sites, respectively, cf. Fig.~\ref{fig:sketch}. In the limit $L_B\to\infty$, the subsystem $A$ reaches a stationary state at large time $t$. Since the model is integrable, it is described by a GGE~\cite{essler2016quench}. Namely, for any observable $\mathcal{O}_A$ supported on $A$, we have $\lim_{t\to\infty}\braket{\Psi_t|\mathcal{O}_A|\Psi_t}={\rm tr}[\rho_{\rm GGE} \mathcal{O}_A]$, where
\begin{equation}\label{eq:GGE}
\rho_{\rm GGE}=\frac{1}{Z}{\rm tr}_B[\exp(-\sum_k \beta_k I_k)].
\end{equation}
Here $\beta_k$ are Lagrange multipliers fixed by the initial state, $I_k$ are integral of motions, $[I_k,H]=0$, while $Z$ is a normalization constant. For the Hamiltonian~\eqref{eq:Ham}, $I_k$ can be identified with the momentum occupation numbers~\cite{calabrese2011quantum_PRL,calabrese2012quantum,calabrese2012quantum_II,fagotti2013reduced} $\hat{n}(k)=\tilde{c}^\dagger_k\tilde{c}_k$, with $\tilde{c}_k$ the Fourier transform of $c_j$. 

In the definition of the GGE, $B$ is traced out. Instead, the PE~\cite{cotler2021emergentquantum,choi2021emergentrandomness} is constructed by measuring and keeping track of the bath. Given a pure state $\ket{\Psi}_{AB}$ on $A$ and $B$ (in our case, the evolved state $\ket{\Psi_t}$), we consider measuring $\hat{n}_j = c^\dagger_j c_j$ at each of the sites in $B$. We denote by $\boldsymbol{z}_B=\{z_{1},\ldots, z_{L_B}\}$ the outcomes [$z_{j}=0,1$], occurring with probability $p(\boldsymbol{z}_B)$. After the measurement, $A$ is in a pure state $\ket{\Psi^\prime(\boldsymbol{z}_B)}_A$, and the PE reads
\begin{equation}\label{eq:projected_ensemble}
	\mathcal{E}^{\rm PE}=\{p(\boldsymbol{z}_B), \ket{\Psi^\prime(\boldsymbol{z}_B)}_A\}\,.
\end{equation}
Averages over this ensemble coincide with expectation values over $\rho_A={\rm tr}_{B}[\ket{\Psi}\bra{\Psi}]$, but the PE contains more information encoded in the higher statistical moments
\begin{equation}\label{eq:higher_moments}
	\rho_{\mathcal{E}}^{(k)}=\sum_{\boldsymbol{z}_{B}} p\left(\boldsymbol{z}_{B}\right)\left(\left|\Psi^\prime\left(\boldsymbol{z}_{B}\right)\right\rangle\left\langle \Psi^\prime \left(\boldsymbol{z}_{B}\right)\right|\right)^{\otimes k}\,.
\end{equation}
Refs.~\cite{cotler2021emergentquantum,choi2021emergentrandomness}, showed that the PE assumes a universal form for chaotic Hamiltonians without conserved quantities, coinciding with a uniform Haar measure over all pure states in $A$. Our goal is to characterize it in the opposite case of an integrable Hamiltonian such as~\eqref{eq:Ham}. 

To simplify the problem, we consider an initial \emph{Gaussian} state~\cite{bravyi2004lagrangian} 
\begin{equation}
	\left|\Psi_{0}\right\rangle_{AB}=\prod_{k=1}^{N}\left[\sum_{j=1}^{L} V_{j k} c_{j}^{\dagger}\right]|\Omega\rangle_{AB}\,,
\end{equation}
where $N$ is the number of particles, $\ket{\Omega}_{AB}$ is the vacuum, while $V$ is a unitary operator. Since the Hamiltonian~\eqref{eq:Ham} is quadratic, $\ket{\Psi_t}_{AB}$ remains Gaussian at all times. In fact, the same is true for the measurement process~\cite{bravyi2004lagrangian}, \emph{i.e.} the projected state $\ket{\Psi^\prime_t}_A$ is also Gaussian. 

This observation allows us to simplify the analysis of the PE. Since Gaussian states satisfy Wick's theorem and $N$ is conserved, all states in the PE~\eqref{eq:projected_ensemble} are completely determined by the corresponding \emph{covariance matrix}
\begin{equation}\label{eq:higher_momentsC}
	[C^\prime_t({\boldsymbol{z}_B})]_{i,j}= \braket{\Psi_t^{\prime}(\boldsymbol{z_B})|c^{\dagger}_ic_j|\Psi_t^{\prime}(\boldsymbol{z_B})}\,,
\end{equation}
with $i,j=1,\ldots L_A$. Thus, higher moments of the PE are encoded in the ensemble $\mathcal{E}^{\rm PE}_C=\{p(\boldsymbol{z}_{B}),C^\prime_t(\boldsymbol{z}_{B})\}$, and the $k$-fold averaged covariance matrices
\begin{equation}\label{eq:higher_moments_covariance}
	C_{\mathcal{E}^{\rm PE}_C}^{(k)}=\sum_{\boldsymbol{z}_{B}} p\left(\boldsymbol{z}_{B}\right) C^\prime_t(\boldsymbol{z}_B)^{\otimes k}\,.
\end{equation}
This is a significant simplification, as the size of covariance matrices scales linearly in the system size.

More importantly, both $p\left(\boldsymbol{z}_{B}\right)$ and $C^\prime_t(\boldsymbol{z}_B)$ can be computed exploiting Gaussianity~\cite{bravyi2004lagrangian}, allowing us to derive exact determinant formulae which can be evaluated efficiently for large system sizes, cf. Appendix~\ref{sec:covariance}. Still, computation of the averages in~\eqref{eq:higher_moments_covariance} remains hard, as the number of terms grows exponentially in $L_B$. To overcome this problem, we have set up a Metropolis Monte Carlo approach, which allows us to sample $p(\boldsymbol{z_B})$ and estimate the averages in~\eqref{eq:higher_moments_covariance}. This method, which takes as an input the covariance matrix of the evolved state, $C_t$, turned out to be very efficient, providing reliable numerical data up to $L_B\simeq 400$ and a relative error of order $10^{-2}$ with $\sim 10^5$ Monte Carlo steps. We provide details of the method in Appendix~\ref{sec:montecarlo}.

\begin{figure}
	\includegraphics[scale=0.4]{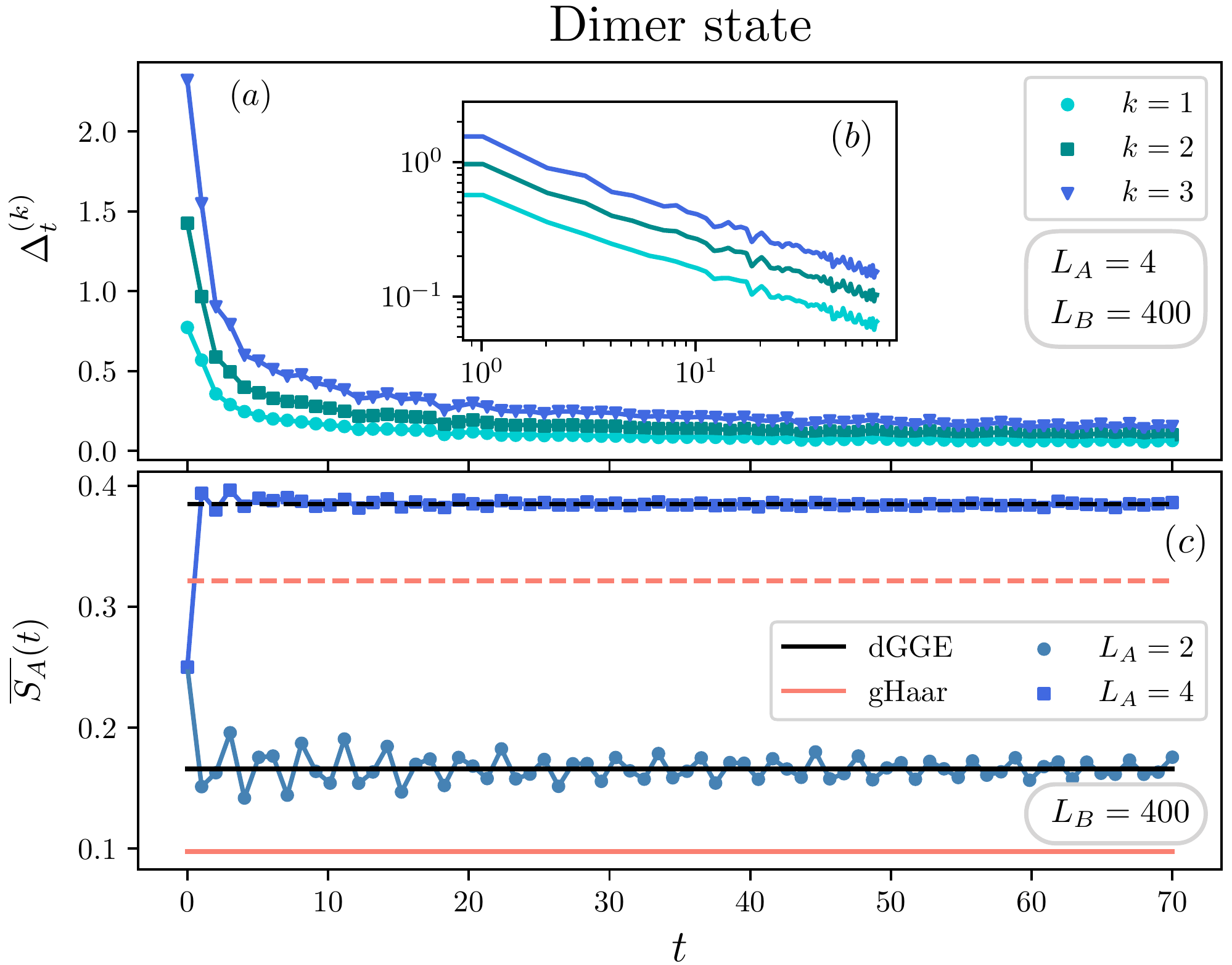}
	\caption{($a$): Difference between the averaged higher moments \eqref{eq:higher_moments_covariance} in the PE $\mathcal{E}^{\rm PE}_C$ and in the dGGE \eqref{eq:EdGGE} (inset $(b)$: log-log plot showing power law decay) from initial state~\eqref{eq:dimer} ($\alpha=e^{i \sqrt{5}}/2$, breaking time-reversal symmetry). $(c)$: space-averaged entanglement entropy $\overline{S_{A}}(t)$, with the corresponding predictions given by the dGGE (horizontal black line) and the infinite temperature ensemble (horizontal red lines). The PE is computed at each time step $\Delta t=1$, using $10^5$ samples. }
	\label{fig:dimer_state}
\end{figure}

\section{The deep GGE} 
\label{sec:deepGGE}

Our goal is to construct a random ensemble, the dGGE, matching the predictions of the PE in the limit $L_B\to\infty$, $t\to\infty$ (in this order). It is useful to imagine that the sites in $B$ are measured sequentially. Each measurement induces a random non-linear transformation of the covariance matrix restricted to $A$. For large $L_B$, it is natural to assume that this causes enough \textit{scrambling} that only a minimal amount of information on the initial state is retained. Thus, it is crucial to identify those features which are preserved by the measurements. Beyond Gaussianity, we know that the dGGE should at least feature complete information on the conserved charges $I_k$, encoded in the Lagrange multipliers $\beta_k$, as it is clear considering the first moment of the PE ensemble [the GGE~\eqref{eq:GGE}]. Following this logic, we propose the \textit{representative-state} approach. 
Considering a pure Gaussian state $\ket{\Phi}_{AB}$, whose conserved charges match those of the initial state $\ket{\Psi_0}_{AB}$, we may define the dGGE as the ensemble obtained by performing projective measurements on subsystem $B$, \emph{i.e.} 
\begin{equation}
\label{eq:EdGGE}
\mathcal{E}^{\rm dGGE}=\{p_{\Phi}(\boldsymbol{z_B}), \ket{\Phi^\prime(\boldsymbol{z}_{B})}_{A} \}\,.
\end{equation}
Here $p_{\Phi}(\boldsymbol{z_B})$ is the probability of obtaining $\boldsymbol{z_B}$ when measuring $\{\hat{n}_i\}_{i\in L_B}$, while $\ket{\Phi^\prime(\boldsymbol{z}_B)}_{A}$ is the post-measurement state. 
To see that $\mathcal{E}^{\rm dGGE}$ correctly reproduces the first moment of the PE~\eqref{eq:projected_ensemble}, we invoke the generalized ETH~\cite{caux2013time,caux2016quench,essler2016quench}, stating $\braket{\Phi| \mathcal{O}_A|\Phi}={\rm tr}[\rho_{\rm GGE} \mathcal{O}_A]$ for all $\mathcal{O}_A$ supported on $A$ and $L_B\to\infty$. 

To test the validity of Eq.~\eqref{eq:EdGGE} beyond the first moment, we perform explicit numerical computations. 
To be concrete, we consider the dimer initial state
\begin{equation}\label{eq:dimer}
	\ket{\Psi_0}=\frac{1}{(1+|\alpha|^2)^{L/4}}(c^\dagger_1+\alpha c^\dagger_2)\cdots (c^\dagger_{L-1} + \alpha c^\dagger_L)|0\rangle\,,
\end{equation}
which is Gaussian and corresponding to a non-trivial GGE for $\alpha\neq 0$, with occupations numbers
\begin{align}\label{eq:nkalpha}
n(k)=\frac{1}{2}+ {\rm Re}\left(\frac{e^{-i k }\alpha}{1+|\alpha|^2}\right)\,.
\end{align}
For finite $L_B$, the PE is sampled using the Monte Carlo approach previously discussed. To sample from the dGGE, we follow two approaches.
The simplest choice for the pure state in Eq.~\eqref{eq:EdGGE} is the \textit{single-eigenstate ensemble}: $\ket{\Phi}_{AB}$ is chosen as a simultaneous eigenstate of all conserved quantities such that the eigenvalues match the expectation values in $\ket{\Psi_0}_{AB}$~\footnote{This definition is inspired by the analysis of Refs.~\cite{cotler2021emergentquantum}, where the equivalence between single-eigenstate ensembles and the PE was established at infinite temperatures}.  In practice, we take an eigenstate of $H$, $\ket{\Phi}=\tilde{c}^\dagger_{k_1}\ldots \tilde{c}^\dagger_{k_{L/2}}\ket{\Omega}$, where $k_{j}$ are drawn randomly according to the distribution function $n(k)$.
A second possibility is to identify $\ket{\Phi}_{AB}$ with a randomly generated correlation matrix $C = U D_{L,N} U^\dag$, where $D_{L,N}$ is a diagonal matrix with $N$ $1$'s and $L-N$ $0$'s. The unitary matrix $U$ is drawn from the following distribution over the appropriate Haar measure, once global symmetries have been taken into account (see below for an example)
\begin{align}\label{eq:PCcan}
    P(U)=  \frac{1}{Z}e^{- \operatorname{Tr}[\Omega F U D_{L,N} U^\dag F^\dag]} \;.
\end{align}
Here $F$ is the Fourier-transform operator mapping the quasimomentum space to the real one. We call this the generalized Haar ensemble: the diagonal matrix $\Omega = \operatorname{diag}(\omega_1,\omega_2,\ldots, \omega_L)$ contains Lagrange multipliers enforcing the constraints $\braket{\Phi_{AB}| \hat{n}(k)|\Phi_{AB}} = n(k)$ [$\omega_k$ should not to be confused with $\beta_k$ appearing in the GGE]. The normalization $Z$ is the Harish-Chandra-Itzykson-Zuber integral~\cite{harish1957differential,itzykson1980planar,mcswiggen2018harish}. Its form is non-trivial but several approximation tools~\cite{8178732, mcswiggen2018harish, PhysRevLett.113.070201} allow determining the functional relation between the $\{\omega_k\}$ and $\{n(k)\}$, as we discuss in Appendix~\ref{sec:gen_haar_ensemble}. 

We sample both the single-eigenstate and the canonical Haar ensembles via the same Monte Carlo approach used for the PE, cf. Appendices~\ref{sec:montecarlo} and~\ref{sec:gen_haar_ensemble}. For sufficiently large $L_B$, we have verified that the two choices for $\ket{\Phi_{AB}}$ give indistinguishable numerical results,
so that in the following we only report data from the single-eigenstate ensemble.

We computed the Frobenious norm~\cite{bhatia2013matrix} of the difference between the $k$-fold averaged covariance matrices~\eqref{eq:higher_moments_covariance} in $\mathcal{E}^{\rm PE}_C$ and $\mathcal{E}^{\rm dGGE}$, denoted by $\Delta^{(k)}_t$. An example of our data is reported in Fig.~\ref{fig:dimer_state}$(a)$, convincingly showing convergence as $t\to\infty$. We see in particular a very clear power-law decay $\Delta^{(k)}_t\sim t^{-1/2}$ independently of $k$. 

As a second non-trivial test, we studied the average of the von Neumann entanglement entropy $S_{A_1}[\boldsymbol{z_B}]=-{\rm tr}\rho_{A_1}(\boldsymbol{z_B})\log \rho_{A_1}(\boldsymbol{z_B})$. Here, $A_1, A_2$ are two subsets of $A$, with $A=A_1\cup A_2$, while $\rho_{A_1}(\boldsymbol{z_B})={\rm tr}_{A_2}[\ket{\Psi^\prime(\boldsymbol{z_B})}\bra{\Psi^\prime(\boldsymbol{z_B})}]$. Since $\ket{\Psi^\prime_t(\boldsymbol{z_B})}$ is Gaussian, $S_{A_1}[\boldsymbol{z_B}]$ can be computed from $C^\prime_t(\boldsymbol{z_B})$~\cite{vidal2003entanglement}, allowing us to sample it via Monte Carlo. Note that $S_{A_1}[\boldsymbol{z_B}]$ involves all higher moments of $C^\prime_t(\boldsymbol{z_B})$, yielding a non-trivial benchmark. In Fig.~\ref{fig:dimer_state}$(b)$, we report our data for the space-averaged entanglement entropy $\overline{S_{A}}(t)$, namely the sum of the values of the bipartite entanglement entropy at each point in $A$, divided by $L_A$. The plot shows very good agreement between the numerical simulation and the result of the ensemble \eqref{eq:PCcan}. We stress that the entanglement entropy under consideration is \textit{not} the one of the GGE, as this quantity is also not a linear functional of the density matrix. Overall, our results consistently support the equivalence between the dGGE and the PE. This is a non-trivial statement, implying that the mere knowledge of the conserved quantities is enough to reconstruct, not only the reduced density matrix, but also all higher moments in \eqref{eq:higher_moments}.

\begin{figure}
	\includegraphics[scale=0.4]{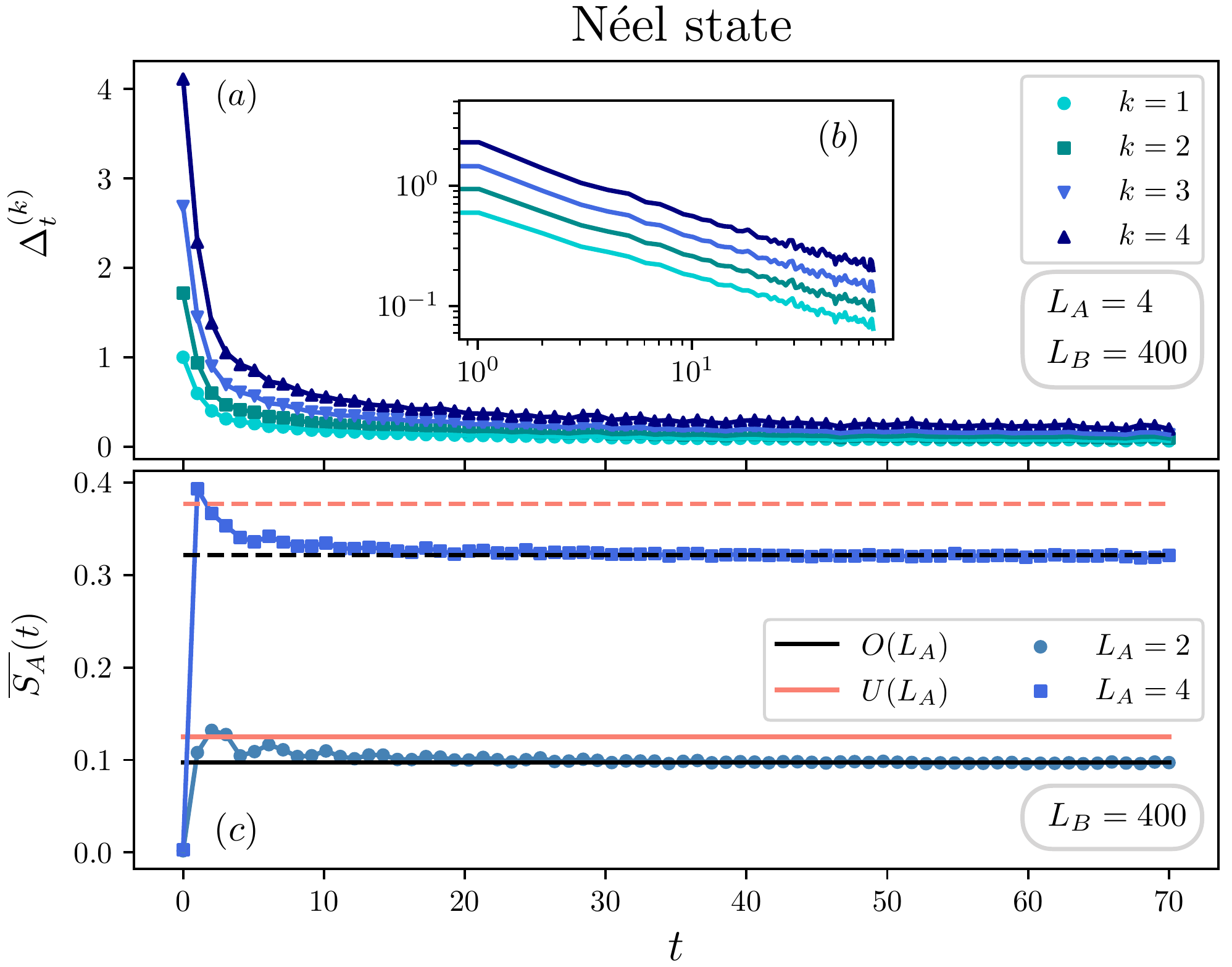}
	\caption{($a$):  Difference between the averaged higher moments \eqref{eq:higher_moments_covariance} in $\mathcal{E}^{\rm PE}_C$ and in the infinite-temperature ensemble~\eqref{eq:ensemble_1}  (inset $(b)$: log-log plot showing power law decay), from initial state~\eqref{eq:dimer} with $\alpha=0$. $(c)$: space-averaged entanglement entropy $\overline{S_{A}}(t)$. We plot the predictions given by the correct ensemble~\eqref{eq:ensemble_1} with orthogonal matrices (horizontal black line) and one where time-reversal symmetry is not correctly enforced, \emph{i.e.} choosing Haar-random unitary matrices (horizontal red lines).}
	\label{fig:neel_state_dyn}
\end{figure}

\section{Infinite-temperature universal ensemble} 
\label{sec:infinite_temp}

The dGGE necessarily contains information on $I_k$, strongly depending on $H$. On the other hand, at infinite-temperatures the GGE loses any information on the latter, suggesting the possibility of a universal description of the PE. We show that this is the case. However, contrary to Refs.~\cite{cotler2021emergentquantum,choi2021emergentrandomness}, the PE takes the form of a uniform measure over the manifold of fermionic Gaussian states. Closely related ensembles appeared in a number of recent works~\cite{liu2018quantum,pengfei2020subsystem,bianchi2021page,bernard2021entanglement,bianchi2021volume,murciano2022symmetry,ulcakar2022tight} and extend the notion of Haar-random states to non-interacting systems.

We focus on the ``N\'eel'' state, obtained by setting $\alpha=0$ in~\eqref{eq:dimer}, corresponding to an infinite-temperature state. From Eq.~\eqref{eq:nkalpha}, one has $n(k) = N/L = 1/2$, so that $\omega_k = 0$ in Eq.~\eqref{eq:PCcan}. It follows that an appropriate correlation matrix for the whole system is obtained by drawing a unitary matrix from the Haar measure. In fact, additional constrains arise due to global symmetries. To elucidate this point, consider the change of basis $R_{\pi/2}=\prod_{j=1}^L e^{i (\pi/2) n_{2j}}$. By inspection, we see that $\ket{\Phi_t}_{AB}=R_{\pi/2}\ket{\Psi_t}_{AB}$ and the projected state $\ket{\Phi^{\prime}_t}_{A}$ are symmetric under time-reversal symmetry $\mathcal{T}$, \emph{i.e.} their wave-function in the canonical basis defined by $c_j^\dagger$ and $\ket{\Omega}_{AB}$ is real. Therefore, the PE can only explore the sector of Gaussian states which is invariant under the joint global symmetry $\mathcal{T}R_{\pi/2}$, and the corresponding ensemble in the space of covariance matrices can be defined 
as $C = R^\dagger_{\pi/2} O D_{L,N} O^\dagger R_{\pi/2}$
where $O$ is drawn from the uniform measure over the orthogonal group $O(L)$~\footnote{Equivalently, one could choose $O$ to be drawn from the special orthogonal group $SO(L)$. However, the two ensemble provide the same physical predictions.}.
Importantly, after the projective measurements, this ensemble can be reduced to one defined only on the sub-system $A$. In particular, the invariance of the Haar measure under left/right multiplication is preserved by the projective measures for all orthogonal transformations restricted to $A$. However, although $\ket{\Psi_t}_{AB}$ has a well-defined particle number, this is not true for the subsystem $A$, and after the measurement it collapses onto a pure state $\ket{\Psi^\prime_t(N_A)}$ with $N_A$ particles, with some probability $p(N_A)$. One can see that the uniform measure with a fixed particle number $N$ for the whole system implies 
that this is only determined by an entropic factor, \emph{i.e.} by the dimensions of the corresponding sector of the Hilbert space, and a random-matrix computation yields $p(N_A)=\binom{L_A}{N_A}2^{-L_A}$, cf. Appendix~\ref{sec:number_dist}. We thus arrive at the following prediction: the PE equals a grand canonical ensemble $\mathcal{E}^{\rm GC}$ over different particle-number sectors each weighted with probability $p(N_A)$. In each sector, it takes the form
\begin{equation}\label{eq:ensemble_1}
	\mathcal{E}^{N_A}=\{C = R^\dagger_{\pi/2} O D_{L_A,N_A} O^\dag R_{\pi/2}\}
\end{equation}
with $O$ uniformly distributed in $O(L_A)$.
$\mathcal{E}^{\rm GC}$ allows us to obtain explicit predictions, by either numerical sampling~\cite{mezzadri2006generate} or analytic formulas derived using the properties of the Haar measure, cf. Appendix~\ref{sec:gen_haar_ensemble}. We have tested it against numerical sampling of the PE. As before, we have studied $\Delta_t^{(k)}$ and the space-averaged entanglement entropy $\overline{S_A}(t)$. An example of our data is reported in Fig.~\ref{fig:neel_state_dyn}, displaying excellent agreement.

Our results show that the infinite-temperature PE is universal even for non-interacting systems, as it only depends on the Gaussianity of the model and on its global symmetries, but not on the details of the Hamiltonian. The same kind of universality was found for instance in Refs.~\cite{vidmar2017entanglement,vidmar2018volume,hackl2019average,lydzba2020eigenstate,lydzba2021entanglement}, studying the averaged entanglement entropy of the eigenstates of quadratic Hamiltonians.

\section{Conclusions}
\label{sec:conclusions}

We have studied the PE emerging at late times after quantum quenches in non-interacting integrable systems. We have characterized it in terms of a random ensemble, the dGGE, constructed out of the initial expectation value of the conserved charges. We have tested our predictions against Monte Carlo sampling of the PE, finding convincing agreement. From the fundamental point of view, our work reveals that, even in non-interacting systems, the PE is largely independent from microscopic details. In particular, at infinite-temperature it coincides with a universal Haar-random ensemble over the set of Gaussian states directly formulated in the subsystem~\cite{liu2018quantum,pengfei2020subsystem,bianchi2021page,bernard2021entanglement,bianchi2021volume,murciano2022symmetry}. This fact could be useful for realizing related ensembles in practice, leveraging the intrinsic randomness of measurements and extending the logic of quantum state designs~\cite{choi2021emergentrandomness,cotler2021emergentquantum}.
For finite temperatures, the existence of a finite correlation length $\xi$ prevents the definition of a post-measurement ensemble expressed uniquely in terms of the charges of $A$. However, this could be possible for $L_A \gg \xi$. We leave this question for future work. Finally, it would be interesting to generalize our study for interacting integrable models where an extensive number of conserved quantities is still present but the Gaussian structure of correlations is lost.

\section*{Acknowledgements} 
J.D.N. acknowledges inspiring discussions with Wen Wei Ho. Some of his work was performed at Aspen Center for Physics, which is supported by National Science Foundation grant PHY-1607611 and at Galileo Galilei Institute during the scientific program ``Randomness, Integrability, and Universality''. This work has been partially funded by the ERC Starting Grant 101042293 (HEPIQ). ADL acknowledges support by the ANR JCJC grant ANR-21-CE47-0003 (TamEnt).


\appendix

\section{Covariance matrix after projective measurement of local densities}
\label{sec:covariance}

In this section we derive the transformation induced on the covariance matrix, defined for any pure state $\ket{\Psi}$ as
\begin{equation}
    C_{ij}=\bra{\Psi}c_{i}^{\dagger}c_{j}\ket{\Psi}.
    \label{correlation}
\end{equation}
We will first consider the effect of measuring the density operator $\hat{n}_\ell = c_\ell^\dag c_\ell$ on a given site $j$ and subsequently the simultaneous effect of many single-site measurements altogether.

\subsection{Single-site measurement}
We foremost observe that after measuring the density operator $\hat{n}_\ell$, two outcomes are possible corresponding to the eigenvalue $z=0$ (empty site) or $z=1$ (occupied site). Denoting as $P_a$ the projector onto the corresponding eigenspace, the post-measurement state can be represented as
\begin{equation}
    \ket{\Psi(z)} = \frac{\Pi_z \ket{\Psi}}{\langle \psi| \Pi_z | \psi \rangle^{1/2} } \;, \qquad 
    \Pi_z = \begin{cases}
    \hat{n}_\ell \;, & z = 1 \\
    1 - \hat{n}_\ell \;, & z = 0
    \end{cases} . 
\end{equation}
Upon an inessential normalisation, we can always represent the projector $\Pi_z \propto \lim_{\mu \to \infty} e^{(z-1/2) \mu \hat{n}_\ell}$, i.e. the exponential of a quadratic operator. This implies that the projective measurement of a local density preserves the Gaussianity of the state. We can thus focus on the transformation induced on the covariance matrix $C_{ij}$. We have
\begin{equation}
    C_{ij}\rightarrow C_{ij}(z) \equiv \frac{\bra{\Psi} \Pi_z c_{i}^{\dagger}c_{j} \Pi_z\ket{\psi}}{\braket{\Psi | \Pi_z | \Psi}} \;, \quad P_z = \braket{\Psi | \Pi_z | \Psi},
    \label{eq:measure}
\end{equation}
where $P_z$ denotes the probability of obtaining the outcome $z$ after the measurement. 
Since after the measurement the state of site $\ell$ factorises, one must have
\begin{equation}
    C_{\ell, \ell}(z) = z \;, \qquad C_{i \ell}(z) = C_{\ell j}(z) = 0 \quad \forall i,j \neq \ell \,.
\end{equation}
We can thus focus on the relevant submatrix $C_{ij}(z)$ with both $i,j \neq \ell$. Let us focus for simplicity on the case $z = 1$. Then, Eq.~\eqref{eq:measure} reduces to (see also e.g.~\cite{PhysRevB.105.094303})
\begin{subequations}
	\label{eq:meassingle1}
\begin{equation}
C_{ij}(z=1) = \frac{\bra{\Psi} c_{i}^{\dagger}c_{j} n_\ell\ket{\psi}}{\braket{\Psi | n_\ell | \Psi}} 
    = C_{ij}-\frac{C_{i\ell}C_{\ell j}}{C_{\ell \ell}} \,,
  \end{equation}
 for  $ i,j \neq \ell $ and
 \begin{equation}
    P_1 = C_{\ell\ell}\,,
\end{equation}
\end{subequations}
where the last equality follows from a simple application of Wick's theorem. The other case $z = 0$ can be obtained 
by a similar calculation or making use of the particle-hole symmetry and leads to
\begin{subequations}\label{eq:meassingle0}
\begin{align}
\forall i,j &\neq \ell \quad C_{ij}(z=0)
= C_{ij}+\frac{C_{i\ell}C_{\ell j}}{1-C_{\ell \ell}} \;, \\  
P_0 &= 1 - C_{\ell\ell}
\;.
\end{align}
\end{subequations}
We can put together (\ref{eq:meassingle1}, \ref{eq:meassingle0}) in the single equation
\begin{subequations}
\label{eq:meassingle}
\begin{align}
\forall i,j& \neq \ell \qquad C_{ij}(z)
= C_{ij}+(-1)^z\frac{C_{i\ell}C_{\ell j}}{P_z} \;, \\  
P_z &= 1 - z - (-1)^z C_{\ell\ell}.
\end{align}
\end{subequations}

\subsection{Measurements on multiple sites}
Now that we understood the effect of measurement on one site, we can generalize it to multiple site measurements. Following the notation of the main text, we assume that the sites undergoing projective measurements of their local densities are all in the spatial region $B$ and we denote as $\boldsymbol{z}_B = \{z_1,\ldots, z_{L_B}\}$, $z_j \in \{0,1\}$ the outcomes of the measurements. We are interested in computing the resulting covariance matrix $C(\boldsymbol{z})_{ij}$ for $i,j \in A$ and the joint probability of all outcomes $P(\boldsymbol{z})$.
\subsubsection{Iterative procedure \label{sec:itproc}}
Since the operators $n_j$ for $j \in B$ all commute to one another, it is clear that measuring all sites in $B$ can be performed as a sequence of single-site measurements with outcomes $\boldsymbol{z}_B$, irrespectively of the order. 
In order to simplify the notation, we assume that the sites are measured from left to right and that the sites in $B$ are the $L_B$ leftmost ones. Let us denote as $\boldsymbol{z}^{(\ell)} = \{z_1,\ldots, z_\ell\}$, i.e. the measurement outcomes of the $\ell$ leftmost sites in $B$. Then, by making use of \eqref{eq:meassingle}, we have
\begin{subequations}
\label{eq:iteproc}
\begin{align}
    P(\boldsymbol{z}^{(\ell+1)}) &= P(\boldsymbol{z}^{(\ell)}) p \;, \\
     p &\equiv (1 - z_{\ell+1} - (-1)^{z_{\ell+1}}  C(\boldsymbol{z}^{(\ell)})_{\ell+1,\ell+1}) , \\
   C_{ij}(\boldsymbol{z}^{(\ell+1)})
=& C_{ij}(\boldsymbol{z}^{(\ell)})\nonumber\\
+&(-1)^{z_{\ell+1}}\frac{C_{i,\ell+1}(\boldsymbol{z}^{(\ell)})C_{\ell+1, j}(\boldsymbol{z}^{(\ell)})}{p} ,
\end{align}
\end{subequations}
and the procedure finishes when $k = L_B$ as $\boldsymbol{z}^{(L_B)} = \boldsymbol{z}_{L_B}$.
\subsubsection{Determinant form}
It is possible to derive a closed determinant form which expresses directly $P(\boldsymbol{z}_{L_B})$ and $C(\boldsymbol{z}_{L_B})$.
In order to do so, we introduce the $L_B \times L_B$ matrix $D(\boldsymbol{z}_{L_B})$ and the $L_B$ dimensional vectors $\vec{c}_j$ as
\begin{widetext}
\begin{equation}
    D(\boldsymbol{z}_{L_B})
    = - \begin{pmatrix}
    (-1)^{z_1} & 0 & \ldots & 0 \\
    0 & (-1)^{z_2} & \ldots & 0 \\
    0 & 0 & \ddots & \vdots  \\
    0 & 0 & \ldots & (-1)^{z_{L_B}}
    \end{pmatrix} \;, \qquad \vec{C}_j = \begin{pmatrix}
    C_{1j}\\
    C_{2j}\\
    \vdots\\
    C_{L_Bj}
    \end{pmatrix},
\end{equation}
also we denote as $C^{(B)}$ the restriction of $C$ to the sites in $B$. Then, we can set
\begin{equation}
    C_{i,j}(\boldsymbol{z}_B) = \frac{1}{P(\boldsymbol{z}_B)} 
    \det_{L_B + 1}  
    \left(\begin{array}{c | c}
    C_{ij} & \vec{C}_i^\dag \cdot D(\boldsymbol{z}_B) \\
    \hline
    \vec{C}_j &  \frac{\mathds{1} - D(\boldsymbol{z}_{L_B})}{2} + C^{(B)}\cdot D(\boldsymbol{z}_{L_B}) 
    \end{array}\right)
    \label{Csigma_elem},
\end{equation}
with the associated probability 
\begin{equation}
    P(\boldsymbol{z}_B) =\det_{L_B}\Big(\frac{\mathds{1}-D({\boldsymbol{z}_B})}{2}+C^{(B)}(\boldsymbol{z}_B)D(\boldsymbol{z}_B) \Big).
    \label{proba}
\end{equation} 
The equivalence between the two procedures can be verified by induction. 

As a benchmark, we check that the sum over all probabilities 
for all possible strings $\boldsymbol{z}_B$
gives 1. Using the variables $\sigma_j= 2 z_j  - 1$ we have 
\begin{equation}
\sum_{\boldsymbol{z}_B} P(\boldsymbol{z}_B) =   [\prod_{j=1}^{L_B} \sum_{\sigma_j = \pm} ]\det [ \delta_{ik} (1- \sigma_k) + C^{(B)}_{ik} \sigma_k]  =[\prod_{j=1}^{L_B} \sum_{\sigma_j = \pm} \sigma_j ] \det [ \delta_{ik} (\sigma_k - 1) + C^{(B)}_{ik}  ] .
\end{equation}
We can expand the determinant as 
\begin{equation}
     \det [ \delta_{ik} (\sigma_k - 1) + C^{(B)}_{ik}  ]  =  \prod_{i=1}^{L_B}[ (\sigma_i - 1)/2 + C^{(B)}_{ii}] + \sum_{P \neq 1}  (-1)^{[P]}\prod_{i=1}^{L_B}  C^{(B)}_{i P(i)}.
\end{equation}
Summing over $[\prod_{j=1}^{L_B} \sum_{\sigma_j = \pm} \sigma_j ] $ we easily notice that only the first term contribute (since the other they miss at least one of the factors $(1-\sigma_i)/2$), and the only term not giving zero is 
$
    [\prod_{j=1}^{L_B} \sum_{\sigma_j = \pm} \sigma_j ] \prod_{i} (\sigma_i - 1)/2 = 1
$. 

In practice, for numerical stability and efficiency, we found it more efficient to perform the measurements over the whole region $B$ using the iterative procedure \eqref{eq:iteproc}.

\vspace{10mm}
\begin{figure}[t!]
	\includegraphics[scale=0.6]{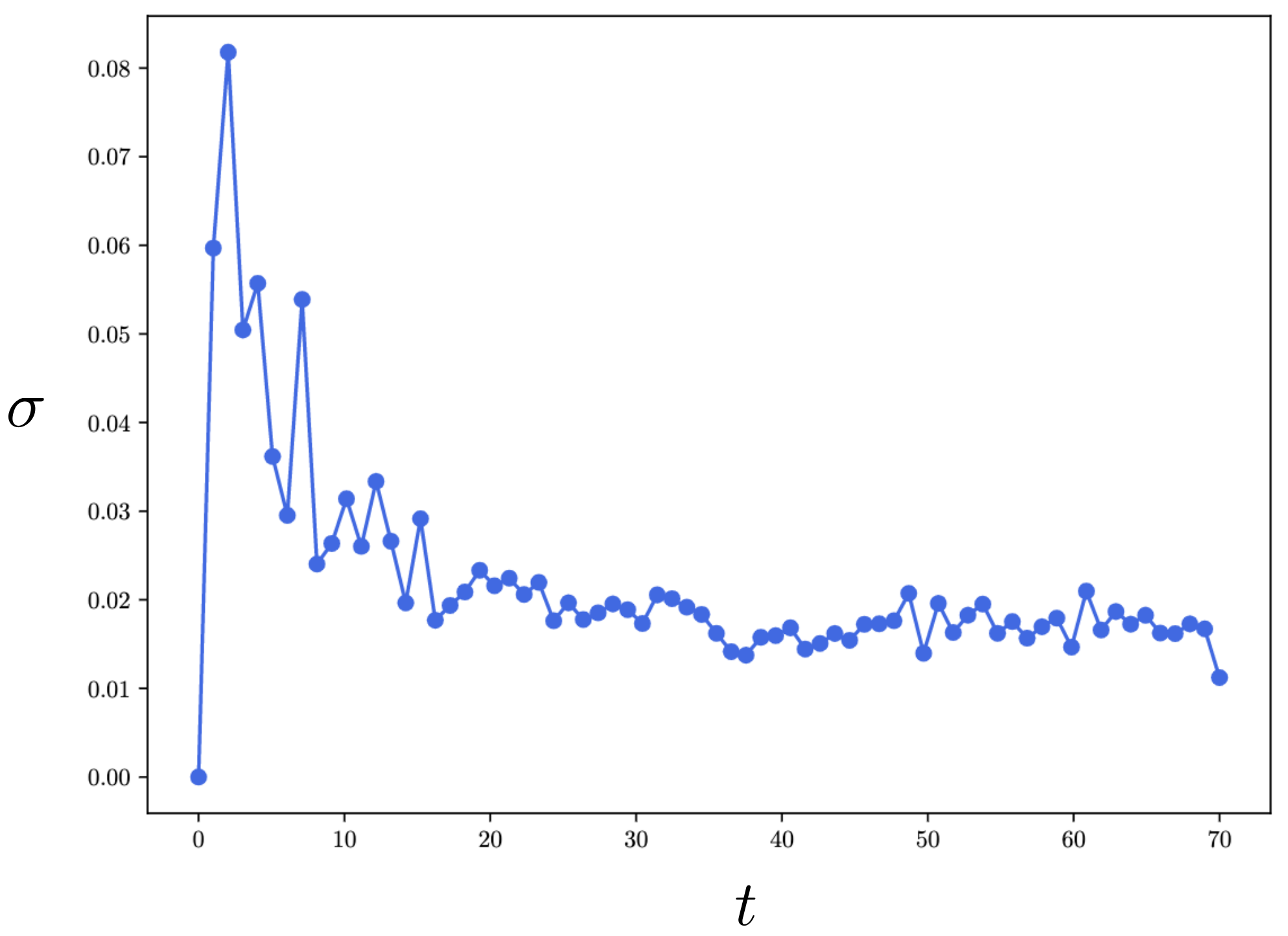}
	\caption{Plot of the relative error $\sigma$ from \eqref{eq:sigma} for the evolution of the total entanglement entropy in $L_A=4$ for the dimer state with $\alpha=0.5 e^{i \sqrt{5}}$.
	}
\label{fig:MC_figure}
\end{figure}

\end{widetext}

\section{Montecarlo sampling of the Projected ensemble}
\label{sec:montecarlo}

In order to compute the PE at any time $t$, we time evolve the correlations matrix using the single particle Hamiltonian $h_{i,j} = \delta_{i,j+1} + \delta_{i,j-1} $ 
\begin{equation}
     C (t) = e^{i h t} C_0 e^{-i h t}
\end{equation}
with the $C_0 = \langle \Psi_0 | c^\dagger_i c_j | \Psi_0 \rangle$ evaluated on the initial state, and at each time step $\Delta t = 1$ we sample the PE by Monte Carlo procedure, namely given the correlation matrix $C$ at time $t$, we start from a random sequence $\boldsymbol{z}^0_B$ of zeros and 1, and we compute $C(\boldsymbol{z}^0_B)$ and $P(\boldsymbol{z}^0_B)$ using the iterative procedure~\eqref{eq:iteproc}. The next Monte Carlo step is to generate a new configuration $\boldsymbol{z}^1_B$ by flipping one 0 or 1 at random within the sequence $\boldsymbol{z}^0_B$ and to compute their ratio of corresponding probabilities 
\begin{equation}
    r = P(\boldsymbol{z}^1_B)/P(\boldsymbol{z}^0_B),
\end{equation}
which is to be compared with a randomly generated real number in the interval $[0,1]$. If the latter is smaller than $r$ the move is accepted and the new correlation matrix is computed as  $C(\boldsymbol{z}^1_B)$, otherwise is rejected and the sequence and the correlation matrix are left unchanged. The algorithm is then iterated on $N_{\rm MC}$ steps where all higher moments of the PE are taken as   
\begin{equation}
\langle 	C^{(k)} \rangle_{\rm MC}= N_{\rm MC}^{-1} \sum_{g=0}^{N_{\rm MC}-1}  C(\boldsymbol{z}^g_B)^{\otimes k}\,.
\end{equation}
In Fig.~\ref{fig:MC_figure} we show the convergence of the von Neumann entropy at different times, by plotting the standard deviation sampled with 40 different realisations of $N_{\rm MC} = 2500$, computed as 
\begin{equation}\label{eq:sigma}
    \sigma = \frac{\sqrt{\langle ( S_A)^2 \rangle_{\rm MC} - \langle ( S_A) \rangle_{\rm MC}}}{\langle ( S_A) \rangle_{\rm MC}},
\end{equation}
 where $S_A$ is the entanglement entropy in the subsystem summed over all sites. The average values are the data reported in the main text.  
The plot shows that expected errors on the Monte Carlo averaging at late times are of order $10^{-2}$.

We note that for the measurements over a set of commuting quantities as we consider here, one can introduce a slightly simpler procedure, which avoids any correlation between configurations produced by the Montecarlo algorithm. In practice, in exactly $L_B$ steps, one generates an entire random sequence $\boldsymbol{z}_B = \{z_1,\ldots, z_{L_B}\}$ with the correct probability: the sites are sequentially measured from left to right, but choosing at each step
\begin{equation}
    z_{\ell} = \begin{cases} 1 & \mbox{with probability } 1 -  C(\boldsymbol{z}^{(\ell-1)})_{\ell,\ell}\\
    0 &  \mbox{with probability } C(\boldsymbol{z}^{(\ell-1)})_{\ell,\ell}
    \end{cases}
\end{equation}
where the correlation matrix $C(\boldsymbol{z}^{(\ell-1)})$ is obtained after the measurement of all sites up to $\ell -1$ (as explained in \ref{sec:itproc}).
See for instance \cite{PhysRevB.105.094303}. Here, we choose to use the more general Montecarlo algorithm explained above.

\section{Generalised Haar ensemble}
\label{sec:gen_haar_ensemble}

As discussed in the main text, a simple way to generate representative states is to sample randomly from an appropriate distribution over the Haar measure, thus enforcing the correct average expectation of the conserved charges. This could be constructed as follows. Because of the conservation of the number of particles, we can always assume that  $C = U D_{L,N} U^\dag$. However, we want to enforce the constraint about the occupation number in the Fourier basis $[F C F^\dag]_{kk} \sim n(k)$, where $F_{kj} = e^{2 \pi i k j/ L}/\sqrt{L}$ performs the change of basis between the momentum and the real space basis. This naturally leads to the \textit{microcanonical Haar} ensemble of covariance matrices
\begin{widetext}
\begin{equation}
\label{eq:PC}
    P(C) = \frac{1}{Z_{\rm MC}}\int_{\rm Haar} dU \, \delta(C - U D(L, N) U^{\dag}) \prod_{k} \delta
\Bigr(\sum_{i=1}^N |[F U]_{k,i}|^2 - n(k)\Bigl),
\end{equation}
where $Z_{\rm MC}$ is the normalization such that $\int P(C) dC = 1$. Note that for an infinite temperature ensemble, $n(k) = N/L$ for any $k$, the delta functions impose no constraints at large $L$ and the matrices $U$ are simply drawn from the uniform distribution over the Haar measure.  
On the contrary, for generic $n(k)$ the deltas forces a bias on the distribution of the matrix $U$. For practical purposes, instead of working with \eqref{eq:PC}, it is better to replace the delta constraint with the \textit{canonical Haar ensemble}, defined by
\begin{align}
    &P(C)  =  \frac{1}{Z[\omega]}\int_{\rm Haar} dU \, \delta(C - U D_{L,N} U^{\dag} ) e^{- \operatorname{Tr}[\Omega F U D_{L,N} U^\dag F^\dag]} = 
    \int_{\rm Haar} d\tU \, \delta(C - F^\dag \tU D_{L,N} \tU^{\dag} F) e^{- \operatorname{Tr}[\Omega \tU D_{L,N} \tU^\dag]} 
    \;, \label{eq:PCcan_SM}
    \\
    &Z[\omega] = 
    \int_{\rm Haar} dU \, e^{- \operatorname{Tr}[\Omega F U D_{L,N} U^\dag F^\dag]} = 
    \int_{\rm Haar} d\tU \, e^{- \operatorname{Tr}[\Omega \tU D_{L,N} \tU^\dag]} ,
    \label{eq:Zmudef}
\end{align}
\end{widetext}
where we made use of the invariance of the Haar measure $\tU \equiv F U$ and introduce the diagonal matrix $\Omega = \operatorname{diag}[\omega_1,\ldots, \omega_L]$ containing the Lagrange multipliers. Their value can be fixed via
\begin{equation}
\label{eq:omegacond}
    \partial_{\omega_{k}} \ln Z[\omega] + n(k) = 0.
\end{equation}
We observe that $Z[\omega_k + c] = e^{-c N} Z[\omega_k]$ for any constant $c$. So, the solution of \eqref{eq:omegacond} are always defined up to a constant, which we fix by imposing the constraint
\begin{equation}
\label{eq:omegaconstr}
    \sum_k \omega_k = 0.
\end{equation}

\subsection{Gaussian approximation}
A simple approximation for the integration over the Haar measure is obtained assuming that all entries of the matrix $U$ are Gaussian distributed for large $L$. In the case of Eq.~\eqref{eq:Zmudef},
for non-zero $\omega$'s, there is a competition between the constraint imposed by unitarity
\begin{equation}
    \sum_{i=1}^L |\tU_{ki}|^2 = 1,
\end{equation}
and the one coming from Eq.~\eqref{eq:omegacond}. To enforce both constraints and a Gaussian distribution of the matrix entries, we rather consider the measure
\begin{equation}
\label{eq:gaussZ}
    Z[\omega, \gamma] = 
    \int d\tU e^{- \sum_{k,i} \gamma_k |\tU_{ki}|^2} e^{- \sum_{k} \sum_{i=1}^N \omega_k |\tU_{k,i}|^2} ,
\end{equation}
and fix the Lagrange multiplier $\gamma$'s and $\omega$'s via the conditions
\begin{align}
    & \partial_{\omega_k} \log Z[\omega, \gamma] + n(k) = 0, \\
    & \partial_{\gamma_k} \log Z[\omega, \gamma] + 1 = 0. 
\end{align}
In the case of the unitary group, $\tU \in U(L)$, the entries are Gaussian complex numbers, so that the integration measure factorises as
\begin{equation}
\label{eq:measU}
    d\tU \to \prod_{k,i} d\Re[\tU_{k,i}]d\Im[\tU_{k,i}],
\end{equation}
leading to the solutions
\begin{equation}
\label{eq:gausssol}
    \gamma_k = L \frac{n(k) - n}{1- n(k)} \;, \qquad \omega_k = - L \frac{n(k) - n}{(1- n(k))n(k)}.
\end{equation}
In the infinite temperature case, the filling function $n(k) = N/L$, all the biases $\omega_k = 0$ consistently with the fact that one can simply sample from the pure Haar distribution.

In principle, one may wonder whether the additional constraint about the normalization of columns should also be imposed, i.e.
\begin{equation}
    \sum_{k=1}^L |\tU_{ki}|^2 = 1 \;, \quad \forall i = 1, \ldots, L.
\end{equation}
It is easy to verify that this constraint is automatically satisfied by the solution above, since
\begin{align}
    &\mathbb{E}\left[\sum_{k=1}^L |\tU_{ki}|^2\right] = 
    \sum_{k} \frac{1}{\gamma_k + \omega_k} \nonumber\\
    &= \frac{1}{L n}\sum_k n(k) = 
    1 \;, i \leq N \\
    &\mathbb{E}\left[\sum_{k=1}^L |\tU_{ki}|^2\right] = 
    \sum_{k} \frac{1}{\gamma_k} \nonumber\\
    &= \frac{1}{L(1-n)}\sum_k (1 - n(k)) = 
    1 \;, i > N,
\end{align}
where $\mathbb{E}[\ldots]$ indicates the average with the measure in Eq.~\eqref{eq:gaussZ}.

A similar approximation can be obtained for other groups, i.e. the orthogonal group, by appropriately changing the integration measure Eq.~\eqref{eq:measU}.
The values of $\omega$'s given in Eq.~\eqref{eq:gausssol} provide a good approximation when $n(k)$ does not vary too much with $k$, i.e. close to infinite temperature, but they are not exact in general. In the next sections, we show different methods to obtain more accurate evaluations.

\subsection{Finite-size evaluation}
The partition sum $Z[\omega]$ can actually be computed explicitly using the Harish-Chandran-Itzykson-Zuber~\cite{harish1957differential,itzykson1980planar} formula. We recall that this formula gives the following integral
\begin{equation}
\label{eq:itzzub}
 \int_{\rm Haar} dU e^{\Tr[A U B U^\dag]} = \frac{\det[e^{\lambda_i^A \lambda_j^B}]}{\Delta[\lambda^{(A)}] \Delta[\lambda^{(B)}]},
\end{equation}
where $\lambda^{(A/B)}_i$ is the spectrum of $A/B$ and we defined the Vandermonde determinant of a set as
\begin{equation}
    \label{eq:vdm}
    \Delta(\lambda) = \prod_{i<j} (\lambda_i - \lambda_j).
\end{equation}
Thus, if we define the diagonal matrix $\Omega = \operatorname{diag}(\omega_k)$, the integral in \eqref{eq:Zmudef} can be evaluated replacing $A \to \Omega$ and $B \to D(L,N)$. However, the matrix $D(L,N)$ has (several) degenerate eigenvalues, since its made of $N$ ones and $L-N$ zeros. In this case, a limit is required to properly compute the rhs of \eqref{eq:itzzub}. To regularise, we set
\begin{widetext}
\begin{equation}
    D_\epsilon(L,N) = \operatorname{diag}(1-\epsilon, 1-2\epsilon, \ldots, 1 - N \epsilon, \epsilon, 2 \epsilon, \ldots, (L-N) \epsilon)  = \operatorname{diag}(d^{(\epsilon)}_i),
\end{equation}
and take the limit $\epsilon \to 0$ at the end. We have clearly
\begin{equation}
    \Delta[d^{(\epsilon)}] = \prod_{i<j} (d^{(\epsilon)}_i-d^{(\epsilon)}_j) =(-1)^{N(N+1)/2}\epsilon^{N(N-1)/2 + (L-N)(L-N-1)/2} G(N+1)G(L-N + 1),
\end{equation}
where $G(x)$ is the Barnes G function. Ignoring numerical factors which are irrelevant in the normalisation, one has in the limit
\begin{equation}
\label{eq:ZHCIZ}
    Z[\omega] = \frac{\det A[\omega]}{\Delta[\omega]},
\end{equation}
where the matrix $A[\omega]$ takes the form
\begin{equation}
\label{eq:matrAdef}
    A[\omega] = 
    \begin{pmatrix}
     1 & \omega_1 & \ldots \omega_1^{N-1} & e^{-\omega_1} & e^{-\omega_1} \omega_1 & \ldots & e^{-\omega_1} \omega_1^{L - N - 1} \\
     1 & \omega_2 & \ldots \omega_2^{N-1} & e^{-\omega_2} & e^{-\omega_2} \omega_2 & \ldots & e^{-\omega_2} \omega_2^{L - N - 1} \\
     \vdots & \vdots &&&&& \vdots \\
     1 & \omega_L & \ldots \omega_L^{N-1} & e^{-\omega_L} & e^{-\omega_L} \omega_L & \ldots & e^{-\omega_L} \omega_L^{L - N - 1},
    \end{pmatrix}
\end{equation}
Although exact, Eq.~\eqref{eq:ZHCIZ} does not allow an efficient evaluation at large $L$, because as already seen in Eq.~\eqref{eq:gausssol}, the $\omega$'s become large with $L$, thus making the exponentials in Eq.~\eqref{eq:matrAdef} hard to evaluate numerically.

\end{widetext}

\subsection{High-temperature expansion}
The large $L$ asymptotics of the HCIZ integral has been investigated in several papers~\cite{8178732, PhysRevLett.113.070201}, see also the introductory review \cite{mcswiggen2018harish}. 
One important result is that it is possible to write down explicitly the ``large temperature'' expansion of Eq.~\eqref{eq:Zmudef} directly in the limit of large $L$ in terms of combinatorial quantities.
First of all, we know already from Eq.~\eqref{eq:gausssol} that at large $L$, the $\omega$'s are going to be scaled linearly with $L$. So we set 
\begin{equation}
    \omega_k = L z(2\pi k/L \equiv p) ,
\end{equation}
where $p$ is the quasiparticle momentum in the thermodynamic limit. With this definition, we can express the moments of the matrix $\Omega$ as
\begin{equation}
     \operatorname{Tr}[\Omega^m] = \sum_{k=1}^L \omega_k^m \to L^{m+1} \int \frac{dp}{2\pi} z(p)^m \equiv L^{m+1} s(m).
\end{equation}
We can now introduce a free energy in the form
\begin{equation}
    F[z] = - \lim_{L\to \infty} \frac{1}{L^2} \ln Z[\omega],
\end{equation}
which is now a functional of $z(p)$. 
We have the expansion in powers of $z$~(see Eq.(2.10) in \cite{mcswiggen2018harish})
\begin{equation}
    F[z] = \sum_{d=1}^\infty \frac{(-1)^d}{d!} 
    \sum_{\alpha, \beta \vdash d} (-1)^{\ell(\alpha) + \ell(\beta)} \vec{H}_0(\alpha,\beta)
    n^{\ell(\beta)}
    \prod_{i=1}^{\ell(\alpha)} s(\alpha_i),
\end{equation}
where the sum runs over the partitions $\alpha, \beta$ of the integer $d$ and $\vec{H}_g(\alpha, \beta)$ are the monotonous Hurwitz numbers associated to the pair of partitions $\alpha, \beta$ computed at genus $g = 0$~\cite{goulden_guay-paquet_novak_2013} and we refer to \cite{mcswiggen2018harish} for their combinatorial definition.
The constraint Eq.~\eqref{eq:omegaconstr} turns into the 
 the equation
\begin{equation}
    \frac{\delta F[z]}{\delta z(k)} = \frac{n(k)}{2\pi}.
\end{equation}
Expanding up to the third order ($d = 3$), one has 
\begin{align}
F[z] &= n s(1) 
-\frac{1}{2} (n-1) n \left(s(1)^2-s(2)\right) \nonumber\\
&+ \frac{1}{3} (n-1) n (2 n-1) \left(2 s(1)^3-3 s(2) s(1)+s(3)\right) \nonumber\\
&+ O(z^4).
\end{align}
Taking the functional derivative and constraining $s(1) = 0$ (consistent with 
\eqref{eq:omegaconstr})
\begin{equation}
    n - (1-n) n z(k)  - (1-n) n (2 n-1) (z(k)^2-s(2)) = n_k,
\end{equation}
where $n = N/L$ is the particle density.
One can easily verify that this is solved up to the order $O(n_k - n)^2$ by
\begin{align}
    z(k) &= \frac{n(k)-n}{(n-1) n}+\frac{(1-2 n) (n(k)-n)^2}{(n-1)^2 n^2}+ A \nonumber\\
    &+ O\left((n(k)-n)^3\right),
\end{align}
where the constant $A$ is put to enforce the constraint $s(1) = 0$. The result is consistent with the Gaussian approximation in \eqref{eq:gausssol}. 

However, at the next order, deviations from the Gaussian approximation appear. As an example, we compute them for the case of the dimer states introduced in Eq.~\eqref{eq:dimer}. Writing for generic $\alpha = \alpha_0 e^{i \theta}$, with $\alpha_0 \in \mathbb{R}$ and $\theta \in [0,2\pi)$, we can write it explicitly as
\begin{equation}
    n_k = \frac 1 2  + \frac{\alpha_0 \cos(k - \theta) }{1 + \alpha_0^2} = \frac 1 2 + \epsilon \cos(k - \theta) \;, \quad\epsilon = \frac{\alpha_0}{1 + \alpha_0^2}.
\end{equation}
After some manipulations, one obtains
\begin{equation}
    z(k) = -4 \epsilon  \cos (k - \theta) -8 \epsilon^3 \cos (k - \theta) \cos (2 (k - \theta)) + O(\epsilon^4),
\end{equation}
which differs from the small $\epsilon$ expansion of the Gaussian approximation \eqref{eq:gausssol}
\begin{equation}
    z_{\rm Gauss}(k) = -4 \epsilon  \cos (k - \theta)-16 \epsilon^3 \cos^3(k - \theta)+O\left(\epsilon^4\right).
\end{equation}
\subsection{Montecarlo sampling from the generalised Haar ensemble}
We now suppose that the values of the $\omega_k$ are known, and we want to sample from the distribution in Eq.~\eqref{eq:PCcan_SM}.
The problem has been also analysed in~\cite{leake2021sampling}, here we discuss a straightforward implementation based on the Metropolis--Hastings algorithm. To do so, we introduce a random walk in the ${\rm SU}(L)$ group. We consider the Markov process at discrete time step $\tau$
\begin{align}
    p_{\tau+1}(\tU) &= p_{\tau}(\tU) \nonumber\\
    + \int_{\rm Haar} &d\tU' \; [P(\tU' \to \tU) p_{\tau}(\tU') -
    P(\tU \to \tU') p_{\tau}(\tU)].
\end{align}
Let us first analyse the simple case $\omega_k = 0$, where one simply needs to sample in from the Haar distribution. Given a certain distribution measure $P(M)$ over hermitian matrices $M$ (that we specify later on), one can set
\begin{equation}
    P_0(\tU \to \tU') = \int dM \; P(M) \delta_{\rm Haar} (\tU' - \tU e^{\imath M} ).
\end{equation}
We stress that the $\delta$ function refers to integration via the Haar measure, i.e., it is defined by
\begin{equation}
\int_{\rm Haar} d\tU \delta_{\rm Haar} (\tU - \tU') f(\tU) = f(\tU').
\end{equation}
It is easy to verify from this definition that $\delta_{\rm Haar} (\tU -\tU'  \tU_0 ) = \delta_{\rm Haar} (\tU' -  \tU \tU_0^\dag) $. Indeed,
\begin{widetext}
\begin{equation}
    \int_{\rm Haar} d\tU' \delta_{\rm Haar} (\tU -  \tU' \tU_0) f(\tU') = 
    \int_{\rm Haar} d\tU'' \delta_{\rm Haar} (\tU - \tU'') f( \tU'' \tU_0^\dag) = f( \tU \tU_0^\dag),
\end{equation}
where in the first equality we changed variable $\tU'' = \tU_0 \tU'$ and we used the invariance of the Haar measure over left multiplication ($d\tU' = d\tU''$). 
We thus see that if we choose $P(M) = P(-M)$, one immediately has
\begin{equation}
    P_0(\tU' \to \tU) = \int dM \; P(M) \delta_{\rm Haar} (\tU - \tU' e^{\imath M}) = 
    \int dM \; P(M) \delta_{\rm Haar} (\tU' -  \tU e^{-\imath M}) = P_0(\tU \to \tU').
\end{equation}
Thus, detail balance is fulfilled with the flat measure $p_n(\tU) \to p_{\rm stat} (\tU) = 1$.

From this construction, it becomes clear how to modify the algorithm to obtain sampling from \eqref{eq:PCcan_SM} via the usual Metropolis-Hastings formula. It is enough to set
\begin{equation}
    P(\tU \to \tU') = P_0(\tU \to \tU') A(\tU,\tU') \;, \qquad A(\tU,\tU') \equiv \min\left[1, e^{\Tr[\Omega \tU' D \tU^{\prime\dag}] - \Tr[\Omega \tU D \tU^\dag]} \right].
\end{equation}
\end{widetext}

In order to do so, it is convenient to specify further the distribution over the hermitian matrices $P(M)$. We choose it as rotations. In other words, 
\begin{itemize}
\item we randomly choose a pair of distinct indices $i,j$ uniformly; 
\item we choose a direction $\alpha = x,y,z$ with equal probabilities $1/3$
\item we choose a random ``angle'' $\phi \in [0,2\pi)$
\item we set
\begin{equation}
    M = \frac 1 2 \phi \sigma_{\alpha}^{(i,j)},
\end{equation}
where $\sigma_{\alpha}^{(i,j)}$ indicates a Pauli matrix in the subspace $(i,j)$ and the identity elsewhere.
\item accept the new unitary $\tU' = \tU e^{\imath M}$ 
\begin{itemize}
    \item with probability $1$ if both $i,j \in \{1,\ldots, N\}$ or $i,j \in \{N+1, \ldots, L\}$, because in both these cases $e^{\imath M} D e^{-\imath M} = D$ as the matrix $D$ restricted to $(i,j)$ is a multiple of the identity;
    \item with probability $1$ if $\alpha = z$, because $|\tU'_{ki}|^2 = |\tU_{ki}|^2$
    and $|\tU'_{kj}|^2 = |\tU_{kj}|^2$, as the transformation only adds a phase;
    \item with probability $p = A(\tU,\tU')$ if $i \in \{1,\ldots N\}$ but $j \in \{N+1,\ldots, L\}$. Note that $A(\tU,\tU')$ can be restricted to the subspace of indices $(i,j)$.
\end{itemize}
\end{itemize}

\subsection{Self-improving Montecarlo method}
The algorithm presented in the previous section assumes that the $\omega_k$ are given and allows sampling from Eq.~\eqref{eq:Zmudef}. In reality, what is given is the density $n(k)$ and the parameters $
\omega_k$ are to be fixed from 
\eqref{eq:omegacond}. 
In practice, starting from some initial estimation for the $
\omega$'s, we can iteratively apply the MC procedure to gradually improve such an estimation.
We introduce the functional
\begin{equation}
\mathcal{F}[\omega] = \frac 12  \sum_k (\partial_{\omega_{k}} \ln Z + n(k))^2.
\end{equation}
The optimal choice of the $\omega$'s lies at the minimum of $\mathcal{F}[\omega]$. We can use gradient descent to improve the current estimation of $\omega_k$:
\begin{align}
\label{eq:omegastepdesc}
    \omega_k^{(n+1)} &= \omega_k^{(n)} - \gamma \frac{\partial{F}}{\partial\omega_k} \nonumber\\
    &= 
    \omega_k^{(n)} - \gamma  \sum_k (\partial_{\omega_{k}} \ln Z + n(k)) \partial_{\omega_{k} \omega_{\ell}} \ln Z \nonumber\\
    &=\omega_k^{(n)} - \gamma  \sum_k (n(k) - \langle \tU_k^2 \rangle) \langle \tU_k^2 \tU_\ell^2 \rangle_c.
\end{align}
where we used $\tU_k^2$ as a shortcut for $\sum_{i=1}^N |\tU_{ki}|^2$.
In practice, we run a few MC steps $N_{\rm it}$ at fixed $\omega$'s, which allow estimating $\langle \tU_k^2 \rangle$ and $\langle \tU_k^2 \tU_\ell^2 \rangle$. Then one can use \eqref{eq:omegastepdesc} to update the values of the $\omega$'s. Note however that the fluctuations due to finite $N_{\rm it}$ prevents from converging to arbitrary accuracy. In practice, after a few iterations, the algorithm cannot improve unless $N_{\rm it}$ is increased. 

\section{Number distribution in Gaussian states}
\label{sec:number_dist}

In this section we prove the formula given in the main text about the distribution of the number of particles in the region $A$.
We assume that the whole system is in a random Gaussian state $\ket{\Psi}_{AB}$ described by the ensemble of covariance matrices 
\begin{equation}
\label{eq:enshaarbeta}
    \mathcal{E}_\beta = \{ C = U D_{L,N} U^\dag\; | \; U \sim \mbox{Haar}_\beta \}
\end{equation}
where the parameter $\beta$ indicates: i) the orthogonal group ($\beta = 1)$, ii) the unitary group $\beta = 2$. We set
\begin{equation}
    p(N_A) = \mathbb{E}[\braket{\Psi| \delta_{\hat{N}_A,N_A} | \Psi}]
\end{equation}
where $\hat{N}_A = \sum_{j \in A} \hat{n}_j$ and the average $\mathbb{E}[\ldots]$ is taken over the ensemble \eqref{eq:enshaarbeta}. 
We claim that the following formula holds
\begin{equation}
\label{eq:pNAcomb}
    p(N_A) = \frac{\binom{L_A}{N_A} \binom{L-L_A}{M-N_A}}{\binom{L}{M}}
\end{equation}
which has a simple combinatorial interpretation as splitting the $M$ particles such that $N_A$ are in $A$ and $M - N_A$ are in $B$. Eq.~\eqref{eq:pNAcomb} can be easily proven for random states over the whole Hilbert space~\cite{bianchi2019typical,murciano2022symmetry}. In the Gaussian case, its proof is less evident. 

We proceed as follows. We first of all introduce the generating function
\begin{align}
\label{eq:gdef}
    g(\lambda)& = \sum_{N_A=0}^{L_A} e^{i \lambda N_A } p(N_A) = 
    \mathbb{E}[\braket{\Psi| e^{i \lambda \hat{N}_A} | \Psi}] \nonumber\\
    &= 
    \mathbb{E}\left[
    \det_{L_A} (1 + (e^{i \lambda} - 1) C^{(A)})
    \right]
\end{align}
where in the last equality we used Wick's theorem to express the expectation value in terms of a determinant of the reduced covariance matrix to the region $A$, i.e. $C^{(A)}_{ij} = C_{ij}$ for $i,j \in A$. 
From this construction, the matrix $C^{(A)}$ is known to be drawn from the $\beta$ Jacobi ensemble~\cite{forrester2010log}. The joint probability distribution function of its eigenvalues $\lambda_1,\ldots, \lambda_{L_A}$ takes the form
\begin{widetext}
\begin{equation}
P(\lambda_1,\ldots, \lambda_{L_A}) = \frac{1}{Z}
\prod_{i=1}^{L_A} \lambda_i^{\beta/2 ( a + 1) - 1}
 (1-\lambda_i)^{\beta/2 (b + 1) - 1}
 \prod_{i<j} |\lambda_i - \lambda_j|^{\beta}
\end{equation}
where the constants $a = M - L_A$ and $b = L - L_A - M$. Setting $z = e^{i\lambda} - 1$, we can thus express 
\begin{equation}
\label{eq:gexp}
    g(\lambda) = \int d\lambda_1\ldots d\lambda_{L_A} 
P(\lambda_1,\ldots, \lambda_{L_A})    \prod_{i} (1 + z \lambda_i) 
= \sum_{k=0}^{L_A} \binom{L_A}{k} z^k Q_k
\end{equation}
where in the last equality we used the symmetry of the integral under the permutation of the eigenvalues. The coefficients $Q_k$ can be expressed in terms of the Aomoto's integral~\cite{enwiki:930603951, 10.1093/qmath/38.4.385} and reads
\begin{multline}
    Q_k =
    \frac{1}{Z}
     \int d\lambda_1\ldots d\lambda_{L_A} 
     \prod_{j=1}^k \lambda_j
\prod_{i=1}^{L_A} \lambda_i^{\beta/2 ( a + 1) - 1}
 (1-\lambda_i)^{\beta/2 (b + 1) - 1}
 \prod_{i<j} |\lambda_i - \lambda_j|^{\beta} = \\=
 \frac{\Gamma (-a-b-2 L_A) \Gamma (-a+k-L_A)}{\Gamma (-a-L_A) \Gamma (-a-b+k-2 L_A)}
\end{multline}
Plugging this last expression in Eq.~\eqref{eq:gexp}, we obtain the final formula
\begin{equation}
\label{eq:ghyp}
    g(\lambda) = \, _2F_1(-a-L_A,-L_A;-a-b-2 L_A;-z) = 
    \, _2F_1(-L_A,-M;-L;-z)
\end{equation}
\end{widetext}
Now, standard manipulations of hypergeometric functions can be used to show that 
Eqs.~(\ref{eq:ghyp},\ref{eq:gdef}) lead to Eq.~\eqref{eq:pNAcomb}, as expected.

\let\oldaddcontentsline\addcontentsline
\renewcommand{\addcontentsline}[3]{}
\bibliography{apssamp}
\let\addcontentsline\oldaddcontentsline

\end{document}